\documentclass[a4paper]{article}

\usepackage[full]{textcomp}
\usepackage{INTERSPEECH2018}
\usepackage{amsmath}
\usepackage{bm}
\usepackage{cite}
\usepackage{color} 
\usepackage{url}
\usepackage{enumitem}
\usepackage{tabularx,booktabs}
\usepackage{newtxtext,newtxmath}
\usepackage{roboto}

\newcolumntype{R}{>{\raggedleft\arraybackslash}X}
\newcolumntype{C}{>{\centering\arraybackslash}X}
\newcolumntype{L}{>{\raggedright\arraybackslash}X}

\title{LibriTTS: A Corpus Derived from LibriSpeech for Text-to-Speech}

\name{Heiga~Zen,~Viet~Dang,~Rob~Clark,~Yu~Zhang,~Ron~J.~Weiss,~Ye~Jia,~Zhifeng~Chen,~Yonghui~Wu}
\address{\GoogleAILogo} 
\email{heigazen@google.com}

\begin{document}

\bmdefine{\bO}{O}
\bmdefine{\bC}{C}
\bmdefine{\bc}{c}
\bmdefine{\bo}{o}
\bmdefine{\bW}{W}
\bmdefine{\bmu}{\mu}
\bmdefine{\bQ}{Q}
\bmdefine{\bq}{q}
\bmdefine{\bw}{w}
\bmdefine{\bU}{U}
\bmdefine{\bL}{L}
\bmdefine{\bu}{u}
\bmdefine{\bZero}{0}
\bmdefine{\bI}{I}
\bmdefine{\bR}{R}
\bmdefine{\bP}{P}
\bmdefine{\br}{r}
\bmdefine{\be}{e}
\bmdefine{\bmm}{m}
\bmdefine{\bsigma}{\sigma}
\bmdefine{\bSigma}{\Sigma}
\bmdefine{\bOmega}{\Omega}
\bmdefine{\bomega}{\omega}
\bmdefine{\bS}{S}
\bmdefine{\bA}{A}
\bmdefine{\bC}{C}
\bmdefine{\bM}{M}
\bmdefine{\bg}{g}
\bmdefine{\bs}{s}
\bmdefine{\bpsi}{\psi}
\bmdefine{\bPsi}{\Psi}
\bmdefine{\bphi}{\phi}
\bmdefine{\bPhi}{\Phi}
\bmdefine{\bPi}{\Pi}
\bmdefine{\bpi}{\pi}
\bmdefine{\bLambda}{\Lambda}
\bmdefine{\blambda}{\lambda}
\bmdefine{\bB}{B}
\bmdefine{\bb}{b}
\bmdefine{\bl}{l}
\bmdefine{\bd}{d}
\bmdefine{\bD}{D}
\bmdefine{\bY}{Y}
\bmdefine{\bG}{G}
\bmdefine{\bp}{p}
\bmdefine{\bxi}{\xi}
\bmdefine{\bmeta}{\eta}
\bmdefine{\bzeta}{\zeta}
\bmdefine{\bk}{k}
\bmdefine{\bK}{K}
\bmdefine{\bF}{F}
\bmdefine{\bv}{v}
\bmdefine{\bX}{X}
\bmdefine{\bx}{x}
\bmdefine{\by}{y}
\bmdefine{\bz}{z}
\bmdefine{\bZ}{Z}
\bmdefine{\bcalX}{\mathcal{X}}
\bmdefine{\bH}{H}
\bmdefine{\bh}{h}
\bmdefine{\bcalH}{\mathcal{H}}
\bmdefine{\bV}{V}
\def\diag{\operatorname{diag}}
\def\idiag{\operatorname{diag}^{-1}}
\def\tr{\operatorname{tr}}
\def\vec{\operatorname{vec}}
\def\Gauss{\mathcal{N}}
\def\Qf{\mathcal{Q}}
\def\calM{\mathcal{M}}
\def\Ind{\mathrm{I}}
\def\Err{\mathcal{E}}
\def\Data{\mathcal{D}}
\def\Loss{\mathcal{L}}

\definecolor {GoogleRed}   {rgb}{0.97265625, 0.00390625, 0.00390625}
\definecolor {GoogleBlue}  {rgb}{0.0078125,  0.3984375,  0.78125}
\definecolor {GoogleYellow}{rgb}{0.9453125,  0.70703125, 0.05859375}
\definecolor {GoogleGreen} {rgb}{0.0,        0.57421875, 0.23046875}
\def\GoogleLogo{\textsf{\textcolor{GoogleBlue}{G}\textcolor{GoogleRed}{o}\textcolor{GoogleYellow}{o}\textcolor{GoogleBlue}{g}\textcolor{GoogleGreen}{l}\textcolor{GoogleRed}{e}}}
\def\GoogleAILogo{\GoogleLogo~\textsf{AI}}

\hyphenation{Libri-TTS}
\hyphenation{Libri-Speech}
\hyphenation{Wave-Net}
\hyphenation{Wave-RNN}

\maketitle
\begin{abstract}
This paper introduces a new speech corpus called ``LibriTTS'' designed for text-to-speech use.
It is derived from the original audio and text materials of the LibriSpeech corpus, which has been used for training and evaluating automatic speech recognition systems.
The new corpus inherits desired properties of the LibriSpeech corpus while addressing a number of issues which make LibriSpeech less than ideal for text-to-speech work.
The released corpus consists of 585 hours of speech data at 24kHz sampling rate from 2,456 speakers and the corresponding texts.
Experimental results show that neural end-to-end TTS models trained from the LibriTTS corpus achieved above 4.0 in mean opinion scores in naturalness in five out of six evaluation speakers.
The corpus is freely available for download from \url{http://www.openslr.org/60/}.
\end{abstract}

\section{Introduction}
The introduction of deep learning-based, neural end-to-end approaches has lowered the barrier to develop high-quality text-to-speech (TTS) systems \cite{Char2Wav,DeepVoice1,Tacotron,Tacotron2}.
With a sufficient amount of studio-quality recorded speech data from a single professional speaker, one can train a generative neural end-to-end TTS model capable of synthesizing speech in a reading style almost as natural as the training speaker \cite{Tacotron2,TransformerTTS}.
Thus, the focus of TTS research is shifting toward more challenging tasks, such as creating
multi-speaker TTS systems \cite{DeepVoice2,DeepVoice3,Tacotron2d}, building neural end-to-end TTS systems from small amounts of data \cite{SemiSupervisedEndToEndTTS}, utilizing a small amount of data for voice adaptation  \cite{Baidu_FewShot_Cloning,Tacotron2d,SampleEfficientTTS}, 
investigating unsupervised prosody and speaking style modeling \cite{TacotronGST,GMVAE_Tacotron}, and building TTS voices from noisy found data \cite{VoiceLoop,GMVAE_Tacotron}.

The LibriSpeech corpus \cite{LibriSpeech} is derived from audiobooks that are part of the LibriVox project \cite{LibriVox}.
There are 982 hours of speech data from 2,484 speakers in this corpus.
It is designed to be reasonably balanced in terms of gender and per-speaker duration.
Furthermore, as it is released under a non-restrictive license, it can be used for both non-commercial and commercial purposes.
Although this corpus was originally designed for automatic speech recognition (ASR) research,  
it has been used in various text-to-speech (TTS) research projects \cite{Tacotron2d,SampleEfficientTTS,DeepVoice3} thanks to its attractive properties, such as a non-restrictive license, a large amount of data, and wide speaker diversity.

However, it also has a number of undesired properties when considering its use for TTS .
The properties which are addressed in this paper are as follows:
\begin{itemize}
  \item \emph{The audio files are at 16 kHz sampling rate};  16 kHz sampling is high enough for the ASR purpose but too low to achieve high quality TTS.  Modern production-quality TTS systems often use 24, 32, 44.1, or 48 kHz sampling rate \cite{SiriTTS,ParallelWaveNet}.
  \item \emph{The speech is split at silence intervals}; the training data speech is split at silences longer than 0.3 seconds.   To learn long-term characteristics of speech such as the sentence-level prosody for given a text, it is necessary to split speech only at sentence breaks.
  \item \emph{All letters are normalized into uppercase, and all punctuation is removed}; capitalization and punctuation marks are useful features to learn prosodic characteristics such as emphasis and the length of pauses.
  \item \emph{The position of sentences within paragraphs is discarded}; to learn inter-sentence prosody it is desirable to access neighbouring sentence text or audio, but this information is missing.
  \item \emph{Some audio files contain significant background noise even within its ``clean'' subsets}; in the LibriSpeech corpus, speakers with low word error rates (WERs) using the Wall Street Journal (WSJ) acoustic model were designated as ``clean''.  Therefore, the ``clean'' subset can contain noisy samples.
\end{itemize}
To address these undesired properties while keeping the desired properties (unrestricted license, large speaker inventories, and gender balance) of the LibriSpeech corpus as much as possible, this paper introduces a new corpus called ``LibriTTS''.
The LibriTTS corpus is derived from the original materials (MP3 from LibriVox and texts from Project Gutenberg) of the LibriSpeech corpus and is distributed under the same non-restrictive license.
It has the same speakers and subset split as the LibriSpeech corpus, offering similar gender balance.
The above-mentioned undesired properties in the LibriSpeech corpus are also addressed:
\begin{itemize}
    \item \emph{The audio files are at 24kHz sampling rate}; as most of the original material is recorded at 44.1 or 32kHz sampling rate, all audio with a sampling rate of less than 24kHz were excluded (two 16kHz files and six 22.05kHz files).
    \item \emph{The speech is split at sentence breaks}; book-level texts are split into sentences using Google's proprietary sentence splitting engine then the audio was split at these sentence boundaries.
    \item \emph{Both original and normalized texts are included}; the text has been normalized using Google's proprietary text normalization engine. 
    \item \emph{Contextual information (e.g., neighbouring sentences) can also be extracted}; additional text files provide easy access to neighbouring sentences. 
    \item \emph{Utterances with significant background noise are excluded}; utterance-level signal-to-noise ratio (SNR) is estimated and used to filter out noisy utterances. 
\end{itemize}

The rest of this paper is organized as follows.
Section \ref{sec:datasets} summarizes existing corpora used for TTS research. 
Section \ref{sec:pipeline} describes the data processing pipeline used to produce the LibriTTS corpus from the original materials.
Section \ref{sec:statistics} presents overall statistics of the corpus.
Section \ref{sec:experiment} shows the experimental results of building TTS models with this corpus.
Concluding remarks are given in the final section.

\begin{table*}[t]
\centering
\caption{\label{tab:datasets}
         \textit{List of publicly available English corpora which are often used in recent TTS research papers and their attributes.}}
\vspace{2mm}
\begin{tabularx}{0.98\textwidth}{c * {6}{C}}
\toprule
                               &           &                           &       & Sampling    & Total    \\
Corpus                         & Domain    & License                   & Hours & rate (kHz) & speakers \\ \midrule
ARCTIC \cite{John_ARCTIC_TECH} & Reading   & BSD-style                 & 7     & 16          & 7 \\  
VCTK \cite{VCTK}               & Reading   & ODC-By v1.0 \cite{ODC_BY} & 44    & 48          & 109 \\
BC2011 \cite{BC2011}           & Reading   & Non-commercial            & 16.6  & 16          & 1 \\
BC2013 \cite{BC2013}           & Audiobook & Non-commercial            & 300   & 44.1        & 1 \\
LJSpeech \cite{LJSpeech}       & Audiobook & CC-0 1.0 \cite{CC0}       & 25    & 22.05       & 1 \\
M-AILABS \cite{M_AILABS_DB}    & Audiobook & BSD-style                 & 75    & 16          & 2 \\
LibriSpeech \cite{LibriSpeech} & Audiobook & CC-BY 4.0 \cite{CC_BY}    & 982   & 16          & 2,484 \\
\textbf{LibriTTS}              & Audiobook & CC-BY 4.0 \cite{CC_BY}    & 586 & 24          & 2,456 \\
\bottomrule
\end{tabularx}%
\end{table*}%

\section{Related work}
\label{sec:datasets}

Having appropriate data facilitates exploration of new tasks and research ideas.
Table~\ref{tab:datasets} lists publicly available English speech corpora which are used in recent TTS research papers.

The CMU ARCTIC corpus \cite{John_ARCTIC_TECH} has been used for many years in statistical parametric speech synthesis research \cite{Zen_SPSS_SPECOM}.
However, it is too small to train neural end-to-end TTS models.
The VCTK corpus \cite{VCTK} is popular for experimenting with multi-speaker TTS \cite{DeepVoice2,VoiceLoop,VoiceLoop2,Tacotron2d,SampleEfficientTTS,Lee_VoiceImitation,Liu_cycle} as it contains studio quality speech data from multiple speakers.
The Blizzard Challenge 2011 (BC2011) \cite{BC2011} corpora provides a relatively large amount of reading speech from a single professional speaker.
It is distributed under a non-commercial license.
The LJspeech \cite{LJSpeech}, M-AILABS \cite{M_AILABS_DB}, and Blizzard Challenge 2013 (BC2013) \cite{BC2013} corpora include tens of hours of audio and text from audiobooks read by single speakers.
The LJspeech and M-AILABS corpora are comprised of non-professional audiobooks from the LibriVox project \cite{LibriVox} and distributed under a non-restrictive license, whereas the BC2013 is read by a professional speaker but distributed under a non-commercial license like BC2011.
As they are audiobook recordings, they contain expressive lines and a wide range of prosodic variation.
They are often used for building single-speaker TTS voices \cite{TacotronGST,TacotronProsodyTransfer,Tacotron_Text,GMVAE_Tacotron,MixRepresentationTTS,Mozilla_TTS}.

\section{The data processing pipeline}
\label{sec:pipeline}

We align the long-form audio recordings with the corresponding texts, and split them into sentence-level segments.  
We also need to exclude utterances with audio/text mismatches which can be caused by inaccuracies in the text, reader-introduced insertions, deletions, substitutions, and transpositions, disfluencies, and text normalization errors.
This section describes the pipeline which we developed to produce the LibriTTS corpus.

\subsection{Text pre-processing}

The first step in the pipeline is to split the book-level text into paragraphs/sentences and perform text normalization.
\begin{enumerate}
\item Book-level texts are first split into paragraphs at consecutive newlines.
\item Each paragraph text is further split into sentences by the proprietary sentence splitter.
\item Non-standard words (e.g., abbreviations, numbers and currency expressions) and semiotic classes \cite{Paul_TTS} (text tokens and token sequences that represent particular entities that are semantically constrained, such as measure phrases, addresses and dates) in the sentences are detected and normalized by a weighted finite state transducer (WFST)-based text normalizer \cite{Kestrel}.
\end{enumerate}

\subsection{Extracting multi-paragraph text}

In the original, unprocessed audio and text materials released from the LibriSpeech site, each audiobook consists of chapter-level audio files (in 128kbps MP3 format) whereas each text is a single file of the entire book.
The second step in the pipeline extracts the partial text corresponding to each chapter-level audio file.
\begin{enumerate}
\item Run ASR (Google Cloud Speech-to-Text API \cite{CloudSpeechAPI}) on the chapter-level audio and get its transcription.
\item Extract chapter-level text from the book-level text by matching the transcription with the book-level text.
\end{enumerate}

\subsection{Align the audio and text}
The third step is to align the audio with the extracted text. 
This is done by the engine used for YouTube's ``auto-sync'' feature.
This feature allows video owners to upload a simple text transcript of the spoken content of a video as an alternative to automatically-created closed captions from ASR \cite{AutoSyncBlog}.
Auto-sync is also used to generate data for ASR acoustic model training \cite{AutoSyncASRU}. Here we used a modified version of Auto-sync to generate the LibriTTS corpus.
The uploaded transcript, containing no timing information, is force-aligned (``auto-sync'ed'') using standard ASR algorithms to generate start and end times of each word \cite{AutoSync_automaton}.

\begin{enumerate}
\item A miniature tri-gram language model (LM) is generated using only the concatenated normalized sentences.
\item The audio is recognized using a decoder graph derived from the mini-LM in combination with a bidirectional long short-term memory (LSTM)-based acoustic model \cite{Hasim_LSTM_Interspeech14}.
\item The decoding result is then edit-distance aligned to the normalized sentences. A sentence is marked as ``aligned'' if all words are matching (with edit-distance of zero).
\item Then start and end times for the sentences are generated from the decoding result. 
\end{enumerate}

\subsection{Post processing}
The final step is post-processing.
We filter out possibly problematic lines based on heuristics found with other corpora and perform normalization.
\begin{itemize}
    \item Filter out sentences with more than 71 words, which are likely to be affected by sentence splitting errors.
    \item Filter out utterances with a large average word duration, which is possibly the result of severe audio/text mismatch.
    \item Normalize the polarity of audio by flipping up-side-down waveforms by ensuring DC offsets are positive. 
    \item Run a silence end-pointer to remove long start and end silences.
    \item Compute SNR of the audio using waveform amplitude distribution analysis (WADA) \cite{WADA_SNR}. Audio with WADA-SNR $<20$dB and $<0$dB are filtered out from the ``clean'' and ``other'' subsets, respectively. 
\end{itemize}
After the post-processing step, pairs of audio and texts (original and normalized) are generated to form the final corpus.
The next section describes the statistics of the generated corpus.

\section{Statistics}
\label{sec:statistics}

\begin{table}[t]
\centering
\caption{\label{tab:summary}
         \textit{Data subsets in LibriTTS.}}
\vspace{2mm}
\begin{tabularx}{0.48\textwidth}{lrrrr}
\toprule
                &        & Female   & Male     & Total \\
Subset          & Hours  & speakers & speakers & speakers \\ \midrule
dev-clean       & 8.97   &  20      & 20       & 40 \\
test-clean      & 8.56   &  19      & 20       & 39 \\
dev-other       & 6.43   &  16      & 17       & 33 \\
test-other      & 6.69   &  17      & 16       & 33 \\
train-clean-100 & 53.78  &  123     & 124      & 247 \\
train-clean-360 & 191.29 &  430     & 474      & 904 \\
train-other-500 & 310.08 &  560     & 600      & 1,160 \\ \midrule
Total           & 585.80 &  1,185   & 1,271    & 2,456 \\
\bottomrule
\end{tabularx}%
\end{table}%

\begin{table}[t]
\centering
\caption{\label{tab:filtering}
         \textit{The numbers of sentences in the original partial text, filtered sentences, and final output.}}
\vspace{2mm}
\begin{tabularx}{0.45\textwidth}{LRR}
\toprule
                  & train-clean-360   & train-other-500 \\ \midrule
Original        & 262,107             & 332,816 \\ \midrule 
Not aligned     &  78,058             &  125,806 \\
Too long        &  1,805              &  1,409 \\
Ave. word dur.  &  26                 &  72 \\
SNR             &  65,718             &  485 \\ \midrule
Final           & 116,500            & 205,044  \\
\bottomrule
\end{tabularx}%
\end{table}%

Table~\ref{tab:summary} provides a summary of all subsets in the LibriTTS corpus.
The amount of yielded audio was significantly lower than that of the LibriSpeech corpus (about 60\%).
This is due to 1) stricter requirement in the alignment step (all words in a sentence must have confidence one)
and 2) SNR-based filtering.
Table~\ref{tab:filtering} shows the numbers of sentences in the original partial text, filtered sentences, and the final output.
It can be seen from the table that about 25\% of sentences were filtered via the SNR threshold (20dB) for the ``clean'' subset.
As the SNR threshold for the ``other'' subset is lower (0dB), the number of filtered sentences is less significant.

\begin{figure}[t]
  \centering
  \includegraphics[width=\linewidth]{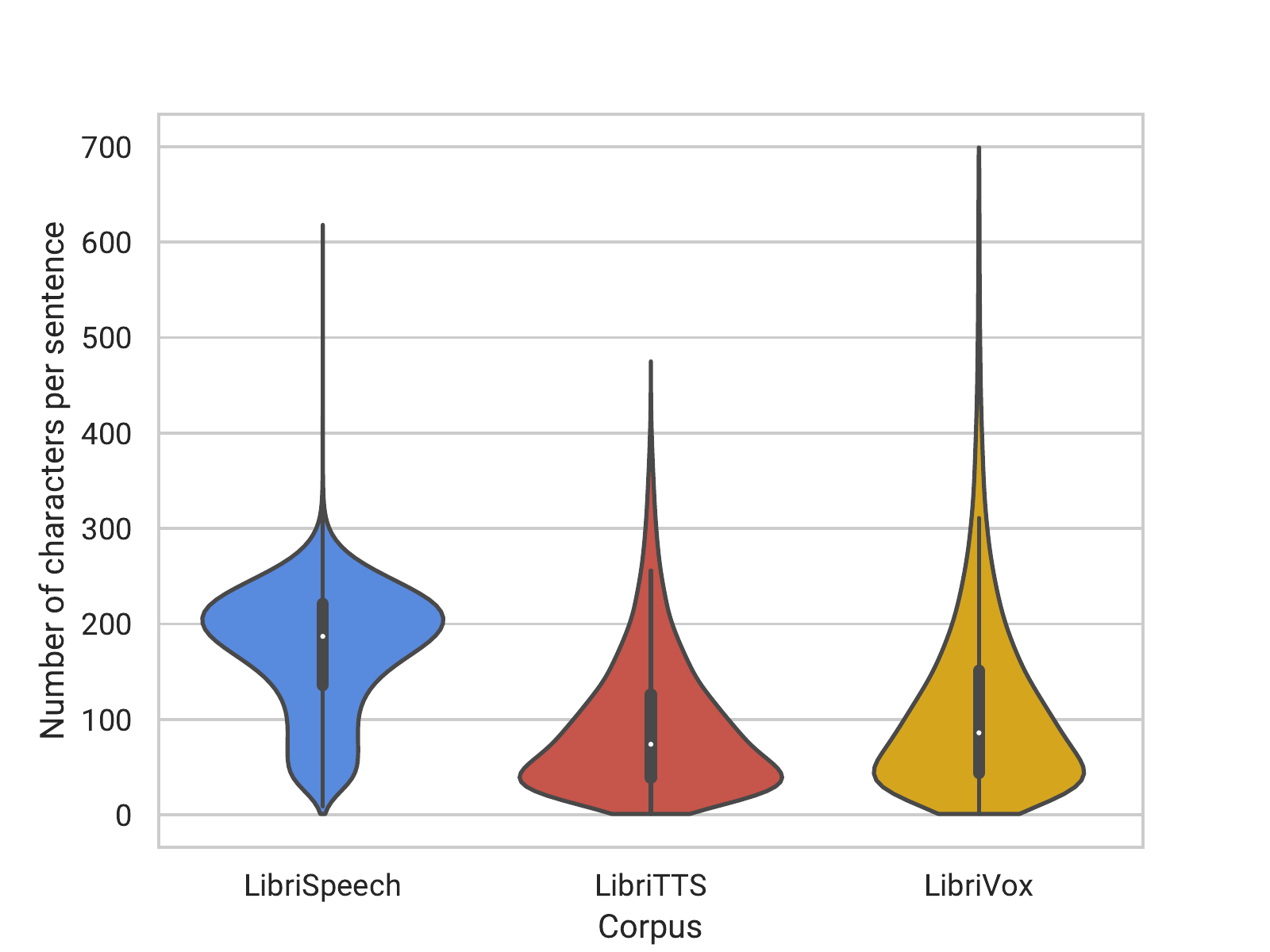}
  \caption{
  Violin plots \cite{ViolinPlot} of the number of characters 
  per sentence on the LibriSpeech and LibriTTS corpora and that of the original LibriVox texts. 
  The thick black bar in the center represents the interquartile range, the thin black line extended from it represents the 95\% confidence intervals, and the white dot is the median.
  The width of the density plot indicates frequency, while each violin is normalized to have the same area.
  }
  \label{fig:histograms_num_chars_words}
\end{figure}

Figure~\ref{fig:histograms_num_chars_words} shows violin plots \cite{ViolinPlot} of the number of character per sentence in the LibriSpeech and LibriTTS corpora and that of the original LibriVox materials.
It can be seen from the figure that the distribution of the sentence length in the LibriTTS corpus is similar to that of the original LibriVox materials, whereas that of the LibriSpeech corpus is significantly different.
Possibly splitting speech at silence intervals rather than sentence boundaries causes this mismatch.
Due to the heuristics to filter-out suspicious long sentences, the distribution for the LibriTTS corpus has shorter tail than that of the LibriVox materials.

\begin{figure}[t]
  \centering
  \includegraphics[width=\linewidth]{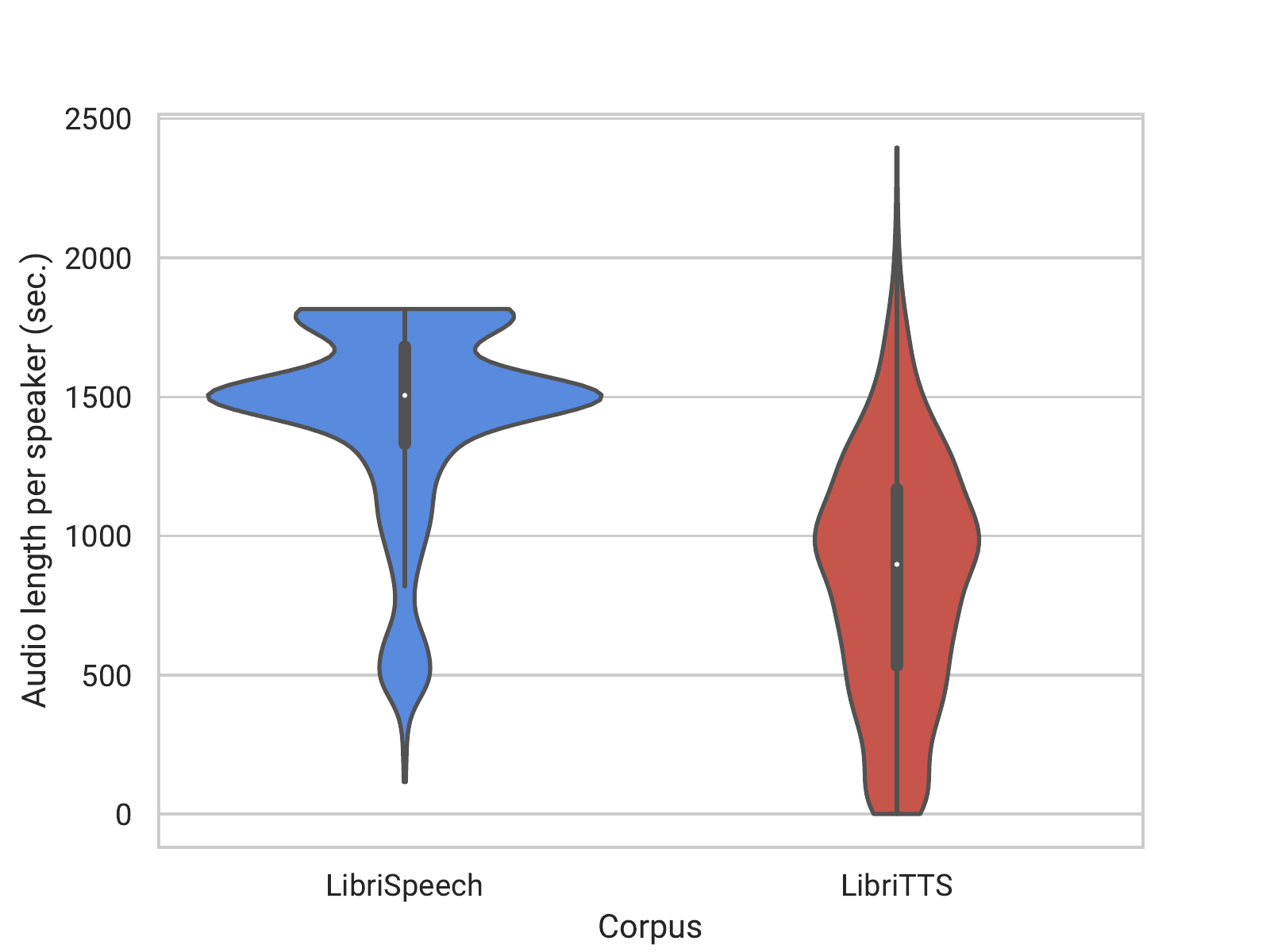}
  \caption{Violin plots \cite{ViolinPlot} of the audio duration per speaker on the LibriTTS and LibriSpeech corpora. 
  }
  \label{fig:histogram_audio_duration}
\end{figure}

One side-effect of the more strict filtering applied in the LibriTTS data creation pipeline is the imbalance in terms of per-speaker duration.
Figure~\ref{fig:histogram_audio_duration} shows violin plots \cite{ViolinPlot} of per-speaker audio duration on the LibriTTS and LibriSpeech corpora.
It can be seen from the figure that the distribution of per-speaker audio duration on the LibriSpeech corpus has a sharp peak at its median (about 1,500 seconds).
On the other hand, that of the LibriTTS corpus has a wider variance and lower median value (about 900 seconds).
Furthermore, its diversity of audio length per speaker is much larger than that of the LibriSpeech corpus.

\begin{table*}[t]
\centering
\caption{\label{tab:MOS}
         \textit{Five-scale subjective mean opinion scores with confidence intervals of the GMVAE-Tacotron models trained from the LibriSpeech and LibriTTS corpora. 
         Total audio duration per speaker in seconds are also included.}}
\vspace{2mm}
\begin{tabularx}{0.85\textwidth}{cccc|ccc}
\toprule
                     & \multicolumn{6}{c}{Speaker} \\ \cmidrule{2-7}
               & \multicolumn{1}{c}{19} & \multicolumn{1}{c}{103} & \multicolumn{1}{c|}{1841} & \multicolumn{1}{c}{204} & \multicolumn{1}{c}{1121} & \multicolumn{1}{c}{5717} \\ \midrule 
LibriSpeech (16kHz)  & 4.03 $\pm$ 0.08 & 4.28 $\pm$ 0.07 & 4.07 $\pm$ 0.07 & 3.98 $\pm$ 0.07 & 3.95 $\pm$ 0.08 & 3.63 $\pm$ 0.08 \\ 
LibriTTS (16kHz)     & 4.19 $\pm$ 0.07 & 4.35 $\pm$ 0.06 & 4.23 $\pm$ 0.07 & 4.01 $\pm$ 0.07 & 3.92 $\pm$ 0.08 & 3.55 $\pm$ 0.09 \\
LibriTTS (24kHz)     & 4.39 $\pm$ 0.06 & 4.51 $\pm$ 0.06 & 4.40 $\pm$ 0.06 & 4.11 $\pm$ 0.08 & 4.05 $\pm$ 0.08 & 3.72 $\pm$ 0.09 \\
Natural (24kHz)      & 4.57 $\pm$ 0.06 & 4.65 $\pm$ 0.09 & 4.57 $\pm$ 0.05 & 4.64 $\pm$ 0.10 & 4.59 $\pm$ 0.08 & 4.46 $\pm$ 0.06 \\ 
\bottomrule
\end{tabularx}%
\end{table*}%

\section{Experiments}
\label{sec:experiment}

This section presents TTS experimental results using models trained from the LibriTTS corpus to give baselines.

\subsection{Experimental conditions}

Gaussian mixture variational auto-encoder (GMVAE)-Tacotron models \cite{GMVAE_Tacotron} were trained from the train-clean subsets of the LibriSpeech and LibriTTS corpora.
The sizes of the latent attribute representation (size of latent vector) and the number of latent attribute classes (the number of mixture components in the Gaussian mixture prior) were 16 and 32, respectively. 
A speaker embedding table was used to give speaker identity conditioning.
Each model was trained for at least 200k steps using the Adam optimizer \cite{Adam}.
Character sequences with punctuation marks from normalized texts were used as inputs of the network.
WaveRNN \cite{WaveRNN}-based neural vocoders at 16kHz and 24kHz sampling rates were trained from the audio in the train-clean subsets of the LibriSpeech and LibriTTS corpora, respectively. 
At synthesis time, first a latent attribute class of the Gaussian mixture prior was randomly selected. Second, a mean vector associated with the selected Gaussian prior was used as the latent attribute representation.
Third, a sequence of log-mel spectrogram was predicted from the normalized input text and the latent attribute representation. 
Finally, a speech waveform was synthesized by driving the WaveRNN neural vocoder given the sequence of log-mel spectrogram.
Please refer to \cite{GMVAE_Tacotron} for details of the hyper-parameters.

Six readers (three female and three male) were selected from the train-clean subsets for evaluation.
The female and male reader IDs were (19, 103, 1841) and (204, 1121, 5717), respectively.
620 evaluation sentences were randomly selected from the test subsets of the corpus.
This set of evaluation sentences is also included in the release.

Quantitative subjective evaluations relied on crowd-sourced mean opinion scores (MOS) rating the naturalness of the synthesized speech by native speakers using headphones.
After listening to each stimulus, a subject was asked to rate the naturalness of the stimulus in a five-point Likert scale score (1: Bad, 2: Poor, 3: Fair, 4: Good, 5: Excellent) in increments of 0.5.
Each sample was rated by a single rater.
To evaluate the effect of sampling rate, we also down-sampled the synthesized speech samples from the LibriTTS model to 16kHz and included them as test stimulus.

\subsection{Results}

Table~\ref{tab:MOS} shows the experimental results.
It can be seen from the table that LibriTTS (24kHz) achieved the best subjective scores with all speakers.
The gaps in MOS between LibriTTS (16kHz) and LibriTTS (24kHz) for female and male speakers were 0.175 and 0.133, respectively.
It clearly shows the benefit of having audio at higher sampling rate.
On the other hand, theose between LibriSpeech (16kHz) and LibriTTS (16kHz) for female and male speakers were 0.15 and -0.03, respectively.
Although this result can suggest that preserving capitalization and punctuation marks were less important, this hypothesis is not fully confirmed as the size of the LibriTTS corpus is about half of that of LibriSpeech (245 hours vs. 460 hours).

It is interesting to note that subjective scores for male speakers were significantly lower than those for female speakers; the gaps between female and male were 0.48 and 0.26 for LibriTTS (24kHz) and LibriSpeech (16kHz), respectively.
It indicates that the current model configuration is sub-optional for male speakers (e.g., filter-bank configuration, modeling dependency of time-domain signals by WaveRNN).
Further experiments are required to fully understand the effect of these configurations.

Finally, there are still significant gap in MOS between natural and synthesized speech (0.16 for female, 0.61 for male).
We need further work to improve the naturalness of synthesized speech on this task.

\section{Conclusions}
This paper introduced the LibriTTS corpus, which was designed for TTS use.
It was derived from the original audio and text materials of the LibriSpeech corpus, by automatically aligning audiobooks and their texts, segmenting them into utterances, and filtering noisy transcripts and audio recordings.
The corpus consists of 585 hours of speech data at 24kHz sampling rate from 2,456 speakers and its corresponding texts.
To our knowledge this is the largest TTS-specific corpus.
We demonstrated that Tacotron models trained from this corpus produced naturally sounding speech.
This corpus is released online for public use;
it is freely available for download from \url{http://www.openslr.org/60/}.
We hope that the release of this corpus accelerates TTS research.

Future work includes evaluating the impacts of speaker imbalance, preserving punctuation marks and capitalization, and the relationship between amount of training data and the naturalness of synthesized speech.
We also plan to expand this corpus by adding more speakers and languages.

\section{Acknowledgements}
The authors would like to thank Hank Liao and Hagen Soltau for their helpful comments about YouTube's auto-sync feature.
We also thank Wei-Ning Hsu, Daisy Stanton, RJ Skerry-Ryan, Yuxuan Wang, and Markus Becker for helpful comments.
We would like to express our gratitude to Guoguo Chen, Sanjeev Khudanpur, Vassil Panayotov, and Daniel Povey for releasing the LibriSpeech corpus, and to the thousands of Project Gutenberg and LibriVox volunteers.

\cleardoublepage

\bibliographystyle{IEEEtran}
\footnotesize
\bibliography{references-short}

\appendix

\normalsize
\section{Character coverage}

\begin{table}[t]
    \centering
    \caption{\label{tab:character_summary}
         \textit{The frequencies of characters in train subsets.}}
    \vspace{2mm}
    \begin{tabular}{cr@{\hspace{3em}}cr}
    \toprule
    character & frequency & character & frequency \\ \midrule
    a & 1,983,462 & A & 54,880 \\
    b & 350,324 & B & 35,627 \\
    c & 593,950 & C & 28,634 \\
    d & 1,114,568 & D & 19,919 \\
    e & 3,218,910 & E & 23,231 \\
    f & 557,647 & F & 16,492\\
    g & 500,616 & G & 17,813 \\
    h & 1,608,244 & H & 48,300 \\
    i & 1,605,001 & I & 134,140 \\
    j & 24,521 & J & 9,581 \\
    k & 204,116 & K & 6,491 \\
    l & 1,024,822 & L & 19,706 \\
    m & 614,732 & M & 31,012 \\
    n & 1,717,592 & N & 23,316 \\
    o & 1,927,370 & O & 21,558 \\
    p & 416,240 & P & 21,389 \\
    q & 24,196 & Q & 1,535 \\
    r & 1,467,483 & R & 16,774 \\
    s & 1,561,623 & S & 46,968 \\
    t & 2,279,182 & T & 85,363 \\
    u & 724,136 & U & 4,944  \\
    v & 237,300 & V & 4,823 \\
    w & 571,524 & W & 38,386  \\
    x & 37,921 & X & 185  \\
    y & 501,661 & Y & 16,753  \\
    z & 13,346 & Z & 628 \\ \relax
    ( & 2,153 & ) & 2,211  \\ \relax
    [ & 169 & ] & 177 \\ \relax
    \{ & 3 & \} & 3 \\ \relax
    '' & 184,459 &  ' & 98,929 \\ \relax
    , & 440,993 & . & 318,448 \\
    ! & 28,275 & ? & 32,663 \\
    - & 38,265 & / & 47 \\
    : & 9,786 & ; & 39,964  \\
    \bottomrule
    \end{tabular}
\end{table}

Table~\ref{tab:character_summary} shows the total numbers of occurrences of characters in all ''train'' subsets. 
It shows that the LibriTTS corpus has good coverage of characters including capitalization and punctuation.

\section{Subjective naturalness ratings}

\begin{figure*}[t]
  \centering
  \includegraphics[width=0.98\textwidth]{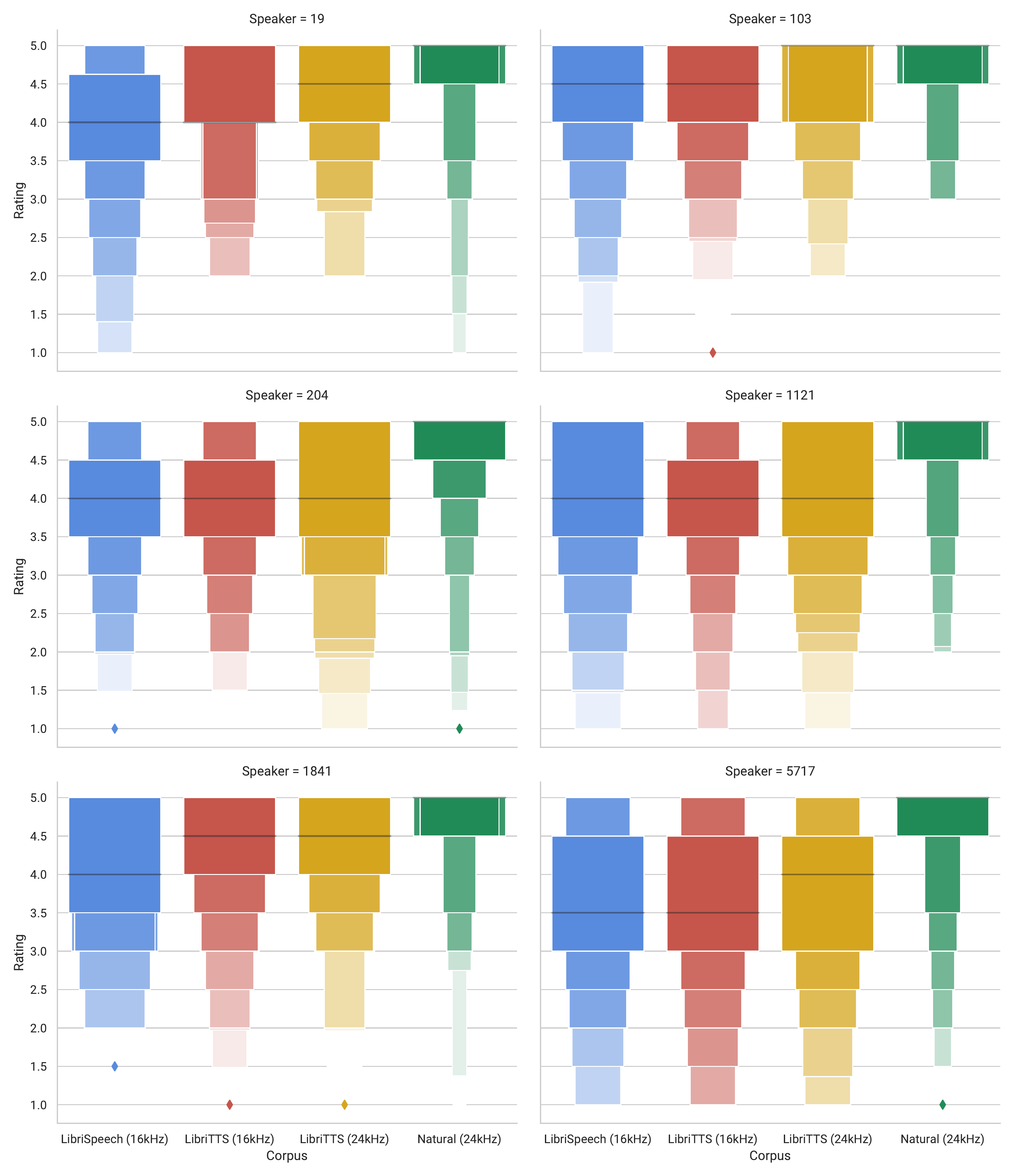}
  \caption{Letter-Value plots \cite{LetterValuePlot} of subjective naturalness ratings for female (above) and male (below) readers.
  }
  \label{fig:mos}
\end{figure*}

Figure~\ref{fig:mos} shows violin plots of subjective naturalness ratings for the experiment conducted in Section~\ref{sec:experiment}.

\end{document}